\documentclass[twocolumn,showpacs,preprintnumbers]{revtex4}
\usepackage{amssymb}
\usepackage{graphicx}
\usepackage{dcolumn}
\usepackage{bm}

\input{tcilatex}

\begin{document}

\preprint{APS/123-QED}
\title{Optimizing the performance of thermionic devices using energy
filtering}
\author{T.E. Humphrey}
\email{tammy.humphrey@unsw.edu.au}
\affiliation{Centre of Excellence for Advanced Silicon Photovoltaics and Photonics,
University of New South Wales 2052, Sydney, Australia.}
\author{H. Linke}
\affiliation{Department of Physics, University of Oregon, Eugene, OR 97403-1274, U.S.A.}
\date{\today}

\begin{abstract}
Conventional thermionic power generators and refrigerators utilize a barrier
in the direction of transport to selectively transmit high-energy electrons.
Here we show that the energy spectrum of electrons transmitted in this way
is not optimal, and we derive the ideal energy spectrum for operation in the
maximum power regime. By using suitable energy filters, such as resonances
in quantum dots, the power of thermionic devices can, in principle, be
improved by an order of magnitude.
\end{abstract}

\maketitle

Thermionic power generators \cite{Hat58, Hou59, Mah98} utilize a temperature
difference between two reservoirs of electrons to transport high-energy
electrons against an electrochemical potential gradient. By increasing the
applied voltage between the reservoirs, the same device can operate in
reverse as a refrigerator \cite{Mah94, Nah94, Ma98C, His01, His03, Chu03},
using the electrochemical potential difference to remove high-energy (`hot')
electrons from the colder reservoir. The required energy selectivity is
conventionally achieved by a barrier between the hot and cold reservoirs
(Fig. \ref{thermionicbarrier}). Thermionic devices may be distinguished from
thermoelectric devices by the use of a barrier which is narrower than the
electron mean free path (ballistic transport) \cite{Mah98}. For the purposes
of this paper, it is important to note that the `energy barriers' used in
conventional devices may more precisely be called `$k_{x}$ barriers', as
they actually constrain the \emph{momentum} of electrons transmitted in
direction of transport so that $k_{x}\geq k_{x}^{\prime }$. While all
electrons with energies less than $E_{B}=\left( \hbar k_{x}^{\prime }\right)
^{2}/2m$ are blocked by such a barrier, \emph{not all electrons with }$E\geq
E_{B}$\emph{\ are transmitted}. In contrast, an `energy filter' may be
understood to be a mechanism which selectively transmits electrons in a
particular range of $E=\left( \hbar k\right) ^{2}/2m$, where $%
k^{2}=k_{x}^{2}+k_{y}^{2}+k_{z}^{2}$. To illustrate this difference, we show
in figure \ref{thermionicbarrier}(b) a Fermi sphere, representing in
momentum space the occupation of states of a free electron gas. The shaded
volume (segment) shows the range of electrons transmitted in the positive $x$
direction by a `$k_{x}$ barrier'. For comparison, the electrons transmitted
by an `energy filter' correspond to a `shell' in $k$ space, as illustrated
in Fig \ref{thermionicfilter} (b).\FRAME{ftbpFU}{8.1012cm}{3.631cm}{0pt}{%
\Qcb{(a) Schematic of a conventional thermionic device, consisting of two
electron reservoirs with different temperatures and electrochemical
potentials. An intervening energy barrier of height $E_{B}=\left( \hbar
k_{x}^{\prime }\right) ^{2}/2m$ constrains the momentum of electrons in the
direction of transport to those with $k_{x}\geq k_{x}^{\prime }$. For
relatively low voltages, there are more high-energy electrons on the hot
side of the barrier, and power is generated by a net electron flow from the
hot the cold reservoirs. If the voltage is increased, the number of
high-energy electrons on the cold side of the barrier increases.At some
voltage the net current direction reverses, and the device cools the cold
reservoir by removing \textquotedblleft hot\textquotedblright\ electrons.
(b) Fermi sphere where the segment for which $k_{x}\geq k_{x}^{\prime }$ has
been shaded.}}{\Qlb{thermionicbarrier}}{thermionicbarrier.gif}{\special%
{language "Scientific Word";type "GRAPHIC";maintain-aspect-ratio
TRUE;display "USEDEF";valid_file "F";width 8.1012cm;height 3.631cm;depth
0pt;original-width 20.1562in;original-height 8.9378in;cropleft "0";croptop
"1";cropright "1";cropbottom "0";filename
'thermionicbarrier.gif';file-properties "XNPEU";}}\qquad

In this paper we show that, from a fundamental point of view, the use of a `$%
k_{x}$ barrier' is not the best possible design for a thermionic device. We
begin by briefly reviewing how energy filtering can be used to achieve
maximum \emph{efficiency }in thermionic devices \cite{Hum02}. Based upon
these results, we find the energy spectrum of electrons that must be
transmitted to achieve maximum power, and so obtain expressions for the
theoretical maximum power of thermionic power generators and refrigerators,
respectively. Finally we compare this ideal case to the spectrum of
electrons actually transmitted by conventional devices using a $k_{x}$
barrier, finding an order of magnitude increase in the maximum power
obtainable from an idealized energy-filtered device compared to a similarly
idealized device which uses a $k_{x}$ barrier.

Hot carrier solar cells \cite{Ros82, Wur97}, quantum dot cryogenic
refrigerators \cite{Edw93, Edw95} and quantum Brownian heat engines \cite%
{Hum02, Hum03} that employ energy filters have been proposed. It has been
shown that ballistic transport of electrons between two reservoirs of free
electron gas is an isentropic process at the energy 
\begin{equation}
E_{0}=\frac{\varepsilon _{C}T_{H}-\varepsilon _{H}T_{C}}{T_{H}-T_{C}}
\label{E01}
\end{equation}%
where the Fermi distributions, $f_{H/C}\left( E_{0}\right) =\left[ 1+\exp
\left( \left[ E_{0}-\varepsilon _{H/C}\right] /kT_{H/C}\right) \right] ^{-1}$%
, in the hot (H) and cold (C)\ reservoirs are equal \cite{Hum02, Hum03}. For
power generation, this energy fulfills the condition $W=\eta Q_{in}$ \cite%
{Ros82, Wur97}, where $W=\left( \varepsilon _{C}-\varepsilon _{H}\right) $
is the work done by each electron transported from the hot to the cold
reservoirs against the electrochemical potential difference, $Q_{in}=\left(
E_{0}-\varepsilon _{H}\right) $ is the heat removed from the hot reservoir
by an electron with energy $E_{0},$ and $\eta =\left( 1-T_{C}/T_{H}\right) $
is the `Carnot factor', the maximum fraction of heat which may be
transformed into useful work by a heat engine working between temperatures $%
T_{H}$ and $T_{C}$. For refrigeration, $E_{0}$ fulfils the condition $%
Q_{out}=W\left[ T_{C}/\left( T_{H}-T_{C}\right) \right] $, where $\left[
T_{C}/\left( T_{H}-T_{C}\right) \right] $ is the coefficient of performance
of a reversible refrigerator and $Q_{out}=\left( E_{0}-\varepsilon
_{C}\right) $ is the heat removed by an electron with energy $E_{0}$ from
the cold reservoir. \FRAME{ftbpFU}{8.1012cm}{3.6728cm}{0pt}{\Qcb{(a) At the
energy $E_{0}$, defined by Eq. \protect\ref{E01}, the Fermi distributions in
the two reservoirs are equal, $f_{C}\left( E_{0}\right) =f_{H}\left(
E_{0}\right) $. The letter G indicates the energy range ($E_{0}<E<\infty $)
for which electrons flow spontaneously from hot to cold, and where power
generation occurs. In the range R ($\protect\varepsilon _{C}<E<E_{0}$)
electrons flow from cold to hot and remove heat from the cold reservoir. (b)
Fermi sphere showing the thin shell of electrons transmitted by an energy
filter.}}{\Qlb{thermionicfilter}}{thermionicfilter.gif}{\special{language
"Scientific Word";type "GRAPHIC";maintain-aspect-ratio TRUE;display
"USEDEF";valid_file "F";width 8.1012cm;height 3.6728cm;depth
0pt;original-width 19.9893in;original-height 8.969in;cropleft "0";croptop
"1";cropright "1";cropbottom "0";filename
'thermionicfilter.gif';file-properties "XNPEU";}}

At $E_{0}$, transport of electrons is reversible and there is no
thermodynamically spontaneous direction for current to flow. A device which
only allowed electrons with this energy to be transmitted would operate with
Carnot efficiency but zero power \cite{Hum02, Hum03}. To find the energy
spectrum of electrons which should be transmitted for maximum power, we note
that power is generated whenever electrons flow from the hot to the cold
reservoir. On the other hand, the cold reservoir is refrigerated when
electrons from above the electrochemical potential in the cold reservoir
flow to the hot reservoir. Over what energy ranges do electrons flow from
the hot to the cold reservoirs and vice-versa?

To proceed, we assume the availability of an idealized energy filter which
transmits all electrons in a desired energy range which arrive at the
interface between reservoirs, and we neglect phonon heat leaks. Using
spherical polar coordinates and working in $k$-space, the particle current
density, $dj_{H}$, of electrons with momentum in the infinitesimal range $dk$
around $k$ arriving at the reservoir interface from the hot reservoir is
given by%
\begin{equation}
dj_{H}\left( k\right) =2\int\limits_{0}^{\pi }\int\limits_{-\pi /2}^{\pi
/2}g\left( \theta ,\phi ,k\right) \nu _{x}\left( \theta ,\phi ,k\right)
f_{H}\left( k\right) d\theta d\phi dk
\end{equation}%
where the density of states is $g=(2\pi )^{-3}k^{2}\sin \theta d\theta d\phi
dk$, the velocity of electrons in the $x$ direction (perpendicular to the
reservoir interface) is $v_{x}=\hbar m^{-1}k\sin \phi \cos \theta $ and the
factor of 2 accounts for electron spin. A similar expression can be written
for the particle current density $dj_{C}$ of electrons arriving at the
interface from the cold reservoir. The net particle current density of
electrons from the hot to the cold reservoirs is then given by $dj=\left(
dj_{H}-dj_{C}\right) $. Evaluating the integral over $\phi $ and $\theta $,
and changing variables to $E=\left( \hbar k\right) ^{2}/2m$, we obtain 
\begin{equation}
dj\left( E\right) =\frac{mE}{2\pi ^{2}\hbar ^{3}}\left[ f_{H}\left( E\right)
-f_{C}\left( E\right) \right] dE  \label{dNdE}
\end{equation}

Assuming that $T_{H}>T_{C}$ and $\varepsilon _{C}>\varepsilon _{H}$, then $%
\left[ f_{H}\left( E\right) -f_{C}\left( E\right) \right] $ is positive for $%
E>E_{0}$, and $dj>0$. This means that electrons in the range $E_{0}<E<\infty 
$ flow from the hot to the cold reservoirs and do work $W=\left( \varepsilon
_{C}-\varepsilon _{H}\right) $ each, while electrons transmitted below $%
E_{0} $ actually reduce the power, each consuming work $W$ as they flow in
the `wrong' direction from the cold to the hot reservoirs. The theoretical
maximum power which can be obtained from a ballistic electron power
generator is therefore 
\begin{equation}
P_{G}=\left( \varepsilon _{C}-\varepsilon _{H}\right)
\int\limits_{E_{0}}^{\infty }dj\left( E\right) .  \label{Ppg}
\end{equation}

Below $E_{0}$, $f_{H}\left( E\right) <f_{C}\left( E\right) $, and $dj\left(
E\right) <0$, so electrons flow from the cold to the hot reservoirs. In
order to refrigerate the cold reservoir transmitted electrons must satisfy $%
E>\varepsilon _{C}$ as well, as the heat change $dQ_{C}$ in the cold
reservoir upon removing an electron with energy $E$ is given by $%
dQ_{C}=E-\varepsilon _{C}.$ Electrons with $E>E_{0}$ flow from hot to cold,
heating the cold reservoir. The theoretical maximum power which can be
obtained from a ballistic electron refrigerator is therefore%
\begin{equation}
P_{R}=-\int\limits_{\varepsilon _{C}}^{E_{0}}\left( E-\varepsilon
_{C}\right) dj\left( E\right) .  \label{Pre}
\end{equation}

We now compare these theoretical limits to the maximum power which may be
obtained from an idealised, conventional thermionic device which utilizes a $%
k_{x}$ barrier. We assume complete transmission for all available electrons
with $k_{x}>k_{x}^{\prime }$ (see Fig. \ref{thermionicbarrier}) and zero
transmission for electrons with $k_{x}<k_{x}^{\prime }$, and find 
\begin{eqnarray}
P_{G}^{Con} &=&\left( \varepsilon _{C}-\varepsilon _{H}\right)
\int\limits_{E_{B}}^{\infty }\left( 1-E_{B}/E\right) dj\left( E\right)
\label{PGcon} \\
P_{R}^{Con} &=&-\int\limits_{E_{B}}^{\infty }\left( 1-E_{B}/E\right) \left(
E-\varepsilon _{C}\right) dj\left( E\right)  \label{PRcon}
\end{eqnarray}%
where $E_{B}=\left( \hbar k_{x}^{\prime }\right) ^{2}/2m$. The
multiplicative term $\left( 1-E_{B}/E\right) $ is a geometrical factor which
occurs due to the fact that only \emph{partial} shells of constant $k$ are
transmitted by devices utilizing a $k_{x}$ barrier.

This factor makes the integrand in Equations \ref{PGcon} and \ref{PRcon}
smaller than that in Equations \ref{Ppg} and \ref{Pre} respectively, for all
electron energies, so $P_{G}^{Con}<P_{G}$ and $P_{R}^{Con}<P_{R}$. For the
refrigeration regime there is an additional source of non-ideality in the
use of a $k_{x}$ barrier which is an important consideration when $%
eV\lesssim kT$. In this case there is substantial occupation of states above 
$E_{0}$, and transmission of electrons with $E>E_{0}$ by a $k_{x}$ barrier
results in a `backcurrent' of hot electrons flowing from the hot to the cold
reservoirs, reducing the refrigerating power below the theoretical maximum,
given by Equation \ref{Pre}. As an illustrative example, for parameter
values of $T_{H}=400\unit{K}$, $T_{C}=300\unit{K}$, $eV=12$m$\unit{eV}$($%
\approx 0.5kT_{C}$) and $m=0.5m_{e}$ (where $m_{e}$ is the mass of a free
electron), and taking $E_{B}=E_{0}$ for power generation, $%
P_{G}/P_{G}^{Con}=17$. Taking $E_{B}=\varepsilon _{C}$ for the refrigeration
case, and $T_{H}=300\unit{K}$, $T_{C}=265\unit{K}$, and the same voltage and
effective mass as before, $P_{R}/P_{R}^{Con}=60$.

The\textit{\ efficiency} of an energy filtered power generator working at
maximum power (Eq. \ref{Ppg}) is also higher than that of a device utilizing
a $k_{x}$ barrier, as a larger proportion of transmitted electrons have
energies close to $E_{0}$ \cite{Hum03}. This increase in efficiency is due
to the fact that while all electrons transmitted from the hot to the cold
reservoirs do work $\varepsilon _{C}-\varepsilon _{H}$, electrons close to $%
E_{0}$ do this work more efficiently than higher energy electrons (which
remove more heat from the hot reservoir than the minimum required by the
second law of thermodynamics). In the refrigeration regime, the efficiency
of an energy filtered device working at maximum power is also higher than
that of a conventional device when $eV\lesssim kT$, due to the supression of
the back-current of high energy electrons. For the device parameters
considered above, the efficiency of the energy filtered power generator is $%
42\%$ of the Carnot limit, while that of the conventional device is $34\%$
of the Carnot limit. The efficiency of the energy filtered electron
refrigerator is $23\%$ of the Carnot limit, compared to $22\%$ of the Carnot
limit for a conventional device. Note that the efficiency of an energy
filtered device can be increased by reducing the range of energies
transmitted to $E_{0}-\delta E<E<E_{0}$, for refrigeration,\ or $%
E_{0}<E<E_{0}+\delta E$, for power generation. When $\delta E\rightarrow 0$,
the Carnot limit is obtained \cite{Hum02, Hum03}.

In principle, suitable energy filtering for electrons could be implemented
via resonant tunnelling through quantum dots \cite{Edw93, Gre00}. A
significant practical loss mechanism for solid-state thermionic devices is
thermal conduction via phonons. In an energy filtered device, this problem
could potentially be tackled via the multilayer approach suggested by Mahan
et al. \cite{Mah98}, to develop a device conceptually similar to that of
Summers and Brennan \cite{Sum86}, or by producing a hybrid
vacuum/solid-state device which incorporated nano-scale voids \cite{Rez03}
together with quantum dots at reservoir interfaces. A hybrid
vacuum/solid-state approach is particularly promising given that the thermal
conductivity of nano-porous silicon ($\sim 0.05\unit{W}\unit{m}^{-1}\unit{K}%
^{-1}$ \cite{Lys00}) is comparable to that of materials such as Bi$_{2}$Te$%
_{3}$ ($\sim 0.07\unit{W}\unit{m}^{-1}\unit{K}^{-1}$ \cite{Nol01}), commonly
used in thermoelectric devices.

\bibliographystyle{apsrev}
\bibliography{acompat,Tammysbib}

\end{document}